\documentclass[11pt,twoside]{article}

\usepackage{asp2006}
\usepackage{epsf}
\usepackage{psfig}
\usepackage{lscape}

\def\mathstacksym#1#2#3#4#5{\def#1{\mathrel{\hbox to 0pt{\lower 
    #5\hbox{#3}\hss} \raise #4\hbox{#2}}}}

\mathstacksym\lta{$<$}{$\sim$}{1.5pt}{3.5pt} 
\mathstacksym\gta{$>$}{$\sim$}{1.5pt}{3.5pt} 
\mathstacksym\lrarrow{$\leftarrow$}{$\rightarrow$}{2pt}{1pt} 
\mathstacksym\lessgreat{$>$}{$<$}{3pt}{3pt} 

\begin{document}

\title{Historical eclipses and the recent solar minimum corona}
\author{P. G. Judge, J. Burkepile, G. de Toma}
{\affil{High Altitude Observatory,
       National Center for Atmospheric Research\footnote{The National %
       Center for Atmospheric Research is sponsored by the %
       National Science Foundation},
       P.O.~Box 3000, Boulder CO~80307-3000, USA}
\author{M. Druckm\"uller}
\affil{Institute of Mathematics,
Faculty of Mechanical Engineering,
Brno University of Technology,
Technicka 2,
616 69 Brno,
Czech Republic
}

\begin{abstract}

We have studied the corona as seen at the eclipses of 1878, 1900, 1901 and
others. These eclipses occurred during extended sunspot minimum conditions. We
compare these data with those of the recent
solar minimum corona, using data from the eclipses of July 22 2009 and 
August 1 2008. An attempt to characterize the global solar
magnetic fields is made. We speculate on the origin of the non-dipolar
structure seen in the 2008 and 2009 eclipse images. 

\end{abstract}


 The SOHO-23 meeting was concerned with understanding if and why the sunspot
 minimum of 2008-2009 is indeed peculiar. We looked at historical eclipse data
 to see how the corona, seen during eclipse, has evolved in time since
 the advent of photography in the mid 1800s. Our goal is to place the 2008
 minimum into the historical record by comparing the images obtained many
 decades ago to those obtained during the recent minimum by one of us (MD).

Since the photographic era began in the mid 1800s, over 100 total solar
eclipses have occurred, a considerable fraction of which were photographed
using telescopes of large focal length. In the late 1960s, Jack Eddy
researched, copied and carefully documented many of the historical eclipse
plates. He compiled data for 20 eclipses, those of 1869, 1878, 1889, 1893,
1898, 1900, 1901, 1905, 1908, 1918, 1922, 1923, 1925, 1932, 1937, 1952, 1962,
1963, 1965, 1966. The copied plates are now stored at the High Altitude
Observatory in Boulder, Colorado. In the 1990s, Eddy's collection was scanned
commercially and the digital images have been archived at HAO.

Solar cycle 11 began in 1867 and as of January 2010
the recent minimum has passed, and cycle 24 has
begun. Of the 12 minima which have occurred since 1869, digitized
data in Eddy's archive exist within 1 year of the minima for the eclipses of
1878, 1900, 1901, 1923, 1965. Digital eclipse data are available from other
sources for the minima of 1986, 1996, and 2008. The following table lists 
circumstances for eclipses during the minima we have examined.
The sunspot data are from the National Geophysical Data Center
(NGDC). 
Cycle lengths were computed from minimum to minimum,
using minima in the mean sunspot number 
from a 13-rotation triangular averaging.

On the basis of sunspot cycle lengths alone, we would have selected,
in order, the years 1901, 1878, 1975 (11.6 yr)  to compare the corona
with that of 2008/2009.  In terms of spotless days we would have
selected 1913, 1901, 1878, 1954. From monthly sunspot numbers we would
have selected 1878, 1954, 1944, 1889, 1922, 1932.  There was no
eclipse in 1913. Patterns in the sunspot behavior alone suggest that the
eclipses of 1900, 1901 and 1878 occurred at phases in
cycles which appear most similar to those of eclipses of the recent minimum.
It would
be helpful to include an analysis of data from the 1954 eclipse, often
cited as a canonical minimum corona \citep[e.g.][]{Billings1966}.
Instead we show data for the 1995 eclipse in Figure 5.

Figure 1 shows the timings of the eclipses on Hathaway's butterfly
diagrams.  The eclipse dates of 1878, 1900, 1901 seem to sample similar phases in their
respective cycles as the 2008 and 2009 eclipses.  
We conclude that the 1878, 1900 and 1901 eclipse
data are indeed appropriate datasets to compare with the recent
minimum corona.  Note that the 1995 eclipse occurs at a period when
significant flux emergence is still occurring.

\begin{table}[!ht]
\caption{Eclipse images examined}
\smallskip
\begin{center}
{\small
\begin{tabular}{llccc}
\tableline
\noalign{\smallskip}
Eclipse date & Location(s) & SSN$^a$ & Spotless days$^a$ & Cycle length$^b$ (yr)\\
\noalign{\smallskip}
\tableline
\noalign{\smallskip}
1878 Jul 29 & Wyoming, Colorado & 3.4 & 280 & 11.7\\
1900 May 28  & Georgia & 9.4 & 158& 11.6\\
1901 May 18 & Sumatra & 2.8 & 287& 11.6\\
1923 Sept 10 & Mexico & 5.8&199& 9.9\\
1965 May 30 & Bellinghausen Island & 15.1 & 70& 10.3\\
1995 Oct 24 & India & 17.5 & 61 & 10.0 \\
2008 Aug 01  & Russia & 2.8 & 266 & 12.6\\
2009 Jul 22 & Marshall Islands & 3.2 & 261 & 12.6\\
 \noalign{\smallskip}
\tableline
\end{tabular}
}
\end{center}
Notes: $^a$These are sunspot numbers and spotless days for the year
during which the eclipse occurred; $^b$This is the length of the
complete cycle
before the date of the eclipse. 
\end{table}

 \setcounter{figure}{0}
 \begin{figure}[!h]
 \plottwo{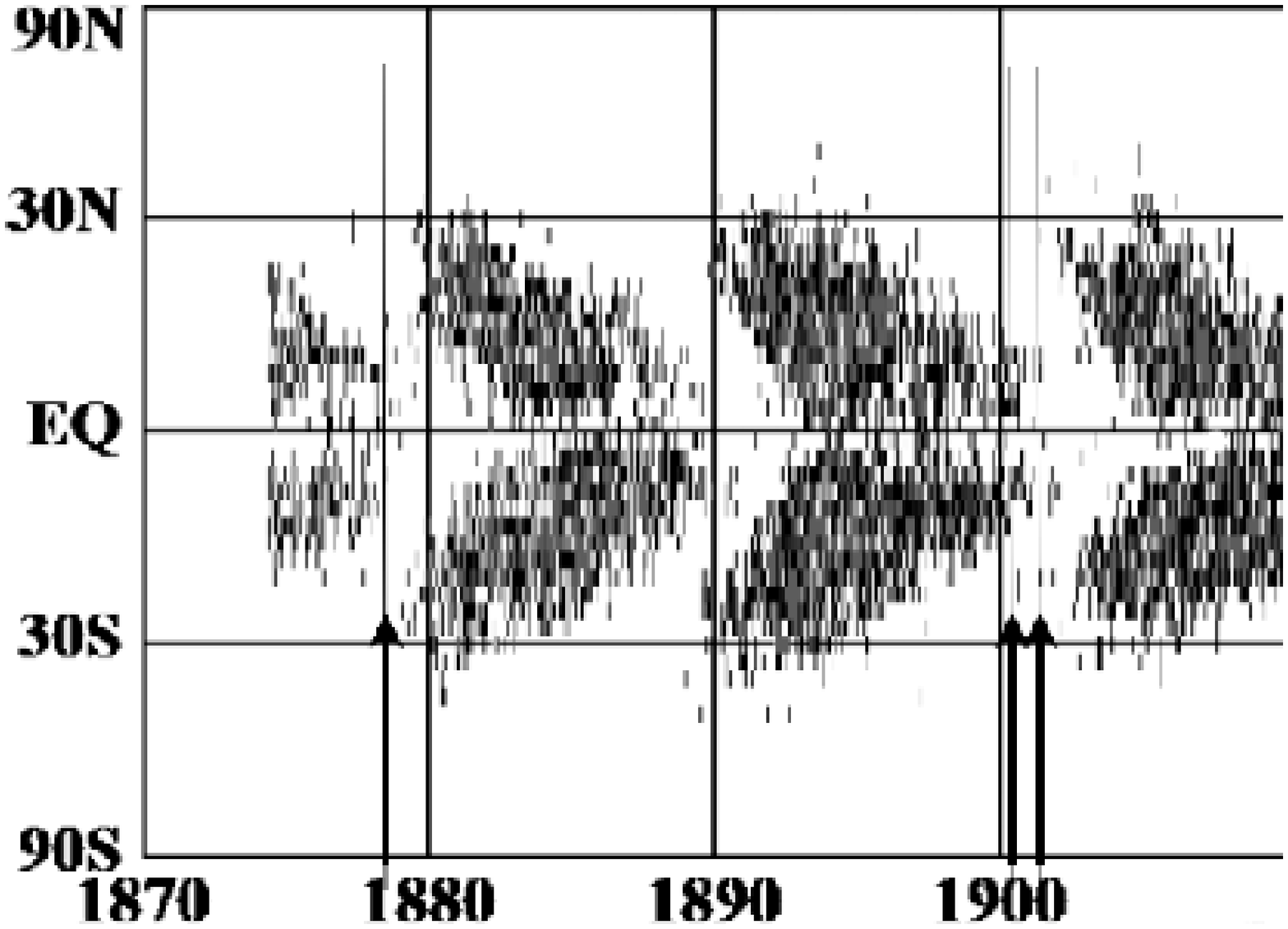}{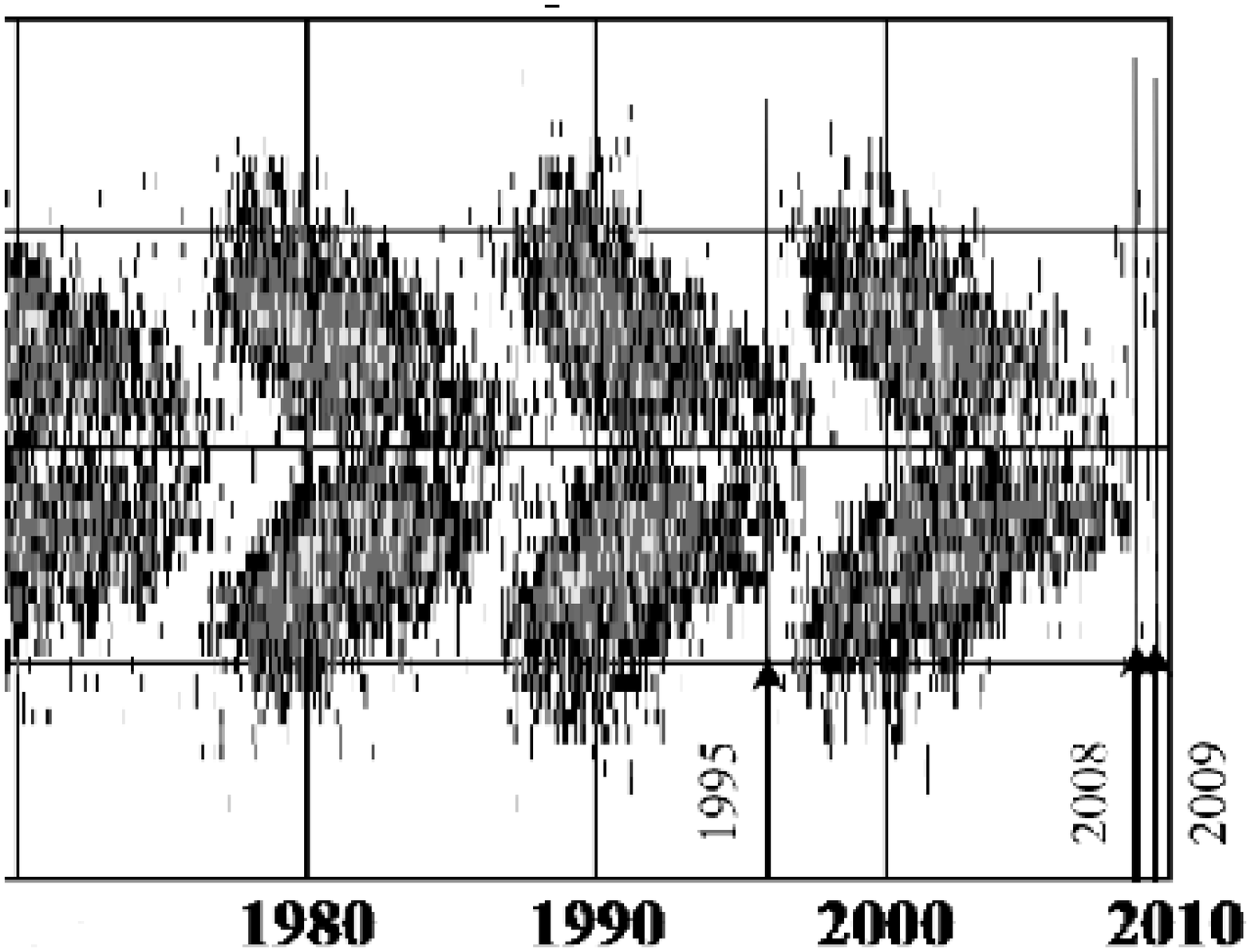}
 \caption{{\itshape Left:\/} Timings of the 1878, 1900 and 1901 plotted on
Hathaway's butterfly diagram.
 {\itshape Right:\/} The same for recent eclipses.}
 \end{figure}

Historical eclipse photographic data are not of a uniform quality.
Furthermore they suffer from limitations inherent in photographic
 techniques and, in a few cases, from damage and/or artifacts.  Before
 1907 plates were sensitive only to blue light \citep{Wallace1907}.
 Figure~2 shows one exposure from the 1878 eclipse obtained by Hall.  In this
 case the dynamic range of the image is limited, and Eddy's plates for
 this eclipse should be re-scanned.  Most of the other
 images in the archive are of a higher quality. 
 \begin{figure}[!h]
\plotfiddle{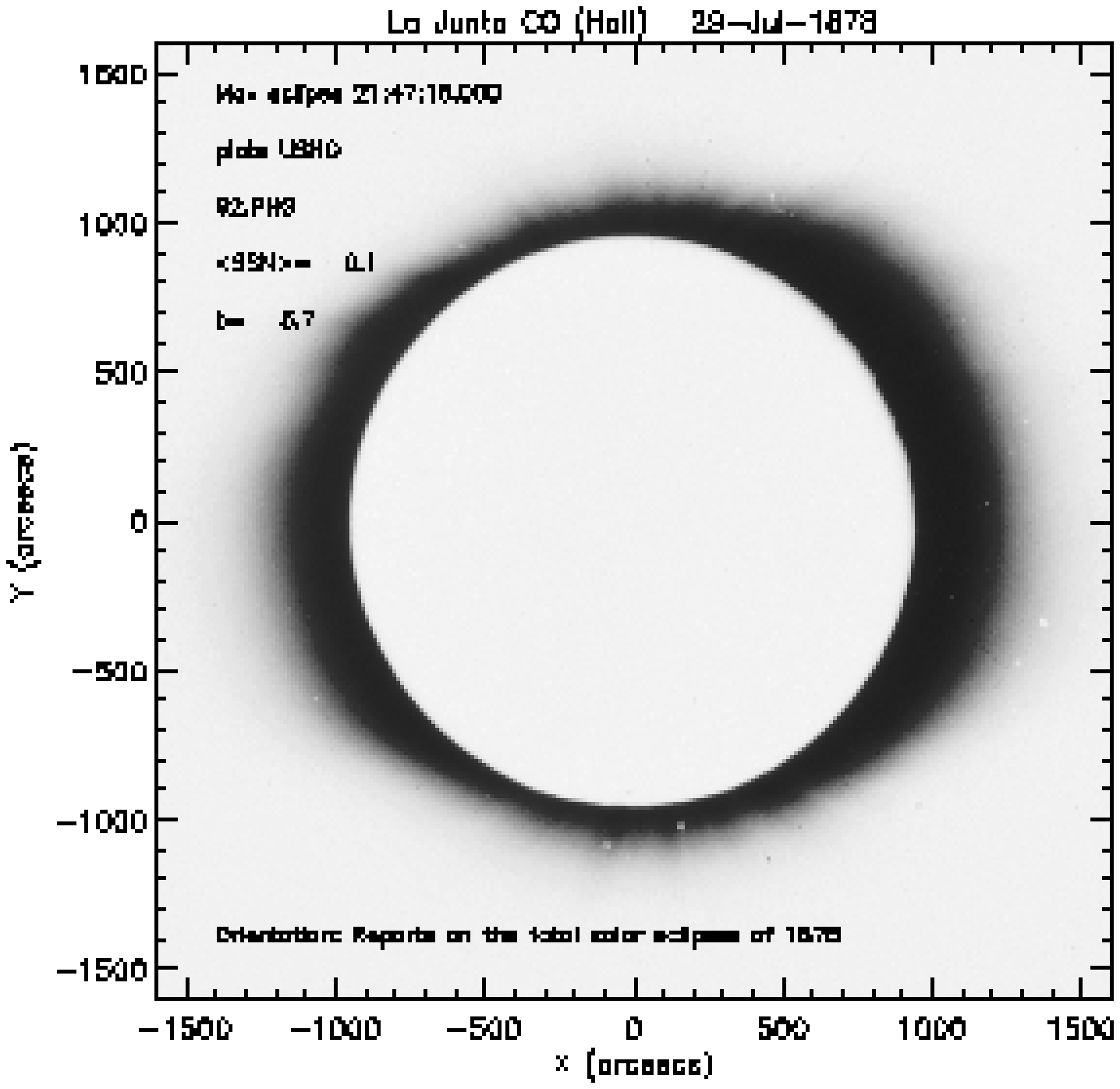}{2.6in}{0.}{50.}{50.}{-220}{-20}
\plotfiddle{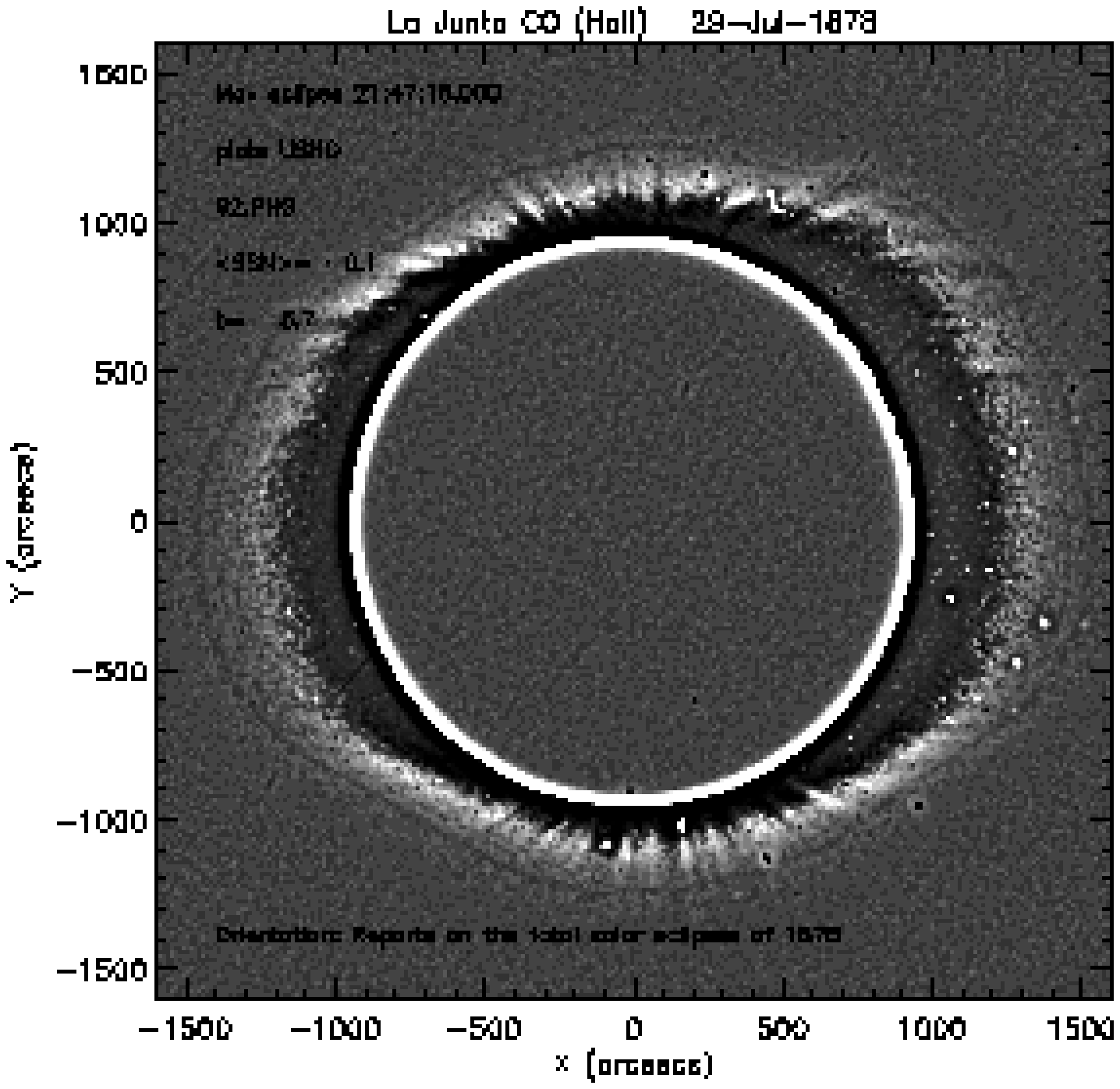}{2.6in}{0.}{50.}{50.}{-10}{180}
\vskip -2.5truein
 \caption{{\itshape Left:\/} One scanned negative plate for the
   eclipse of 1878.  Note the limited dynamic range of the image.
 {\itshape Right:\/} Edge-enhanced version of the same image highlighting
polar plume structure and some plate defects.}
 \end{figure}
The HAO eclipse archive contains data which, as shown in the figure,
are rotated and scaled to the solar orientation (N upwards, E left) 
and apparent angular
size.  The rotations are accurate to about 1 degree and have been done
by eye only, and the scales are set to a fixed angular diameter for
convenience.

We have
 processed some of the data following
 \citet{Druckmueller+Rusin+Minarovjech2006,Druckmueller2009}.  The
processing has two major components- co-alignment and filtering.
Co-alignment is done using a 
modified phase correlation method, based on Fourier transforms, 
enabling the 
translation, rotation, and scaling factors to be found for two
images. 
Pairs of images with different exposure
times and different brightness scales (including non-linear cases like
the photographic records) are treated. 
Adaptive filters inspired
by human vision are applied to sets of eclipse plates. The resulting
co-added and filtered images approximate
what is seen by the human eye during an eclipse, and have minimal
artifacts.   They represent considerable improvements over
unsharp masking methods.  In the case of the 1878 eclipse, the scanned
data
are of insufficient quality to apply these algorithms.  In this
particular
case the sketches made by observers are better suited to examine the 
morphology of the corona  \citep{USNO1890}.  But for other data the
processing is effective.  Figure~3 shows the results applied to the
eclipse of 1901. 

 \begin{figure}[!h]
\plotfiddle{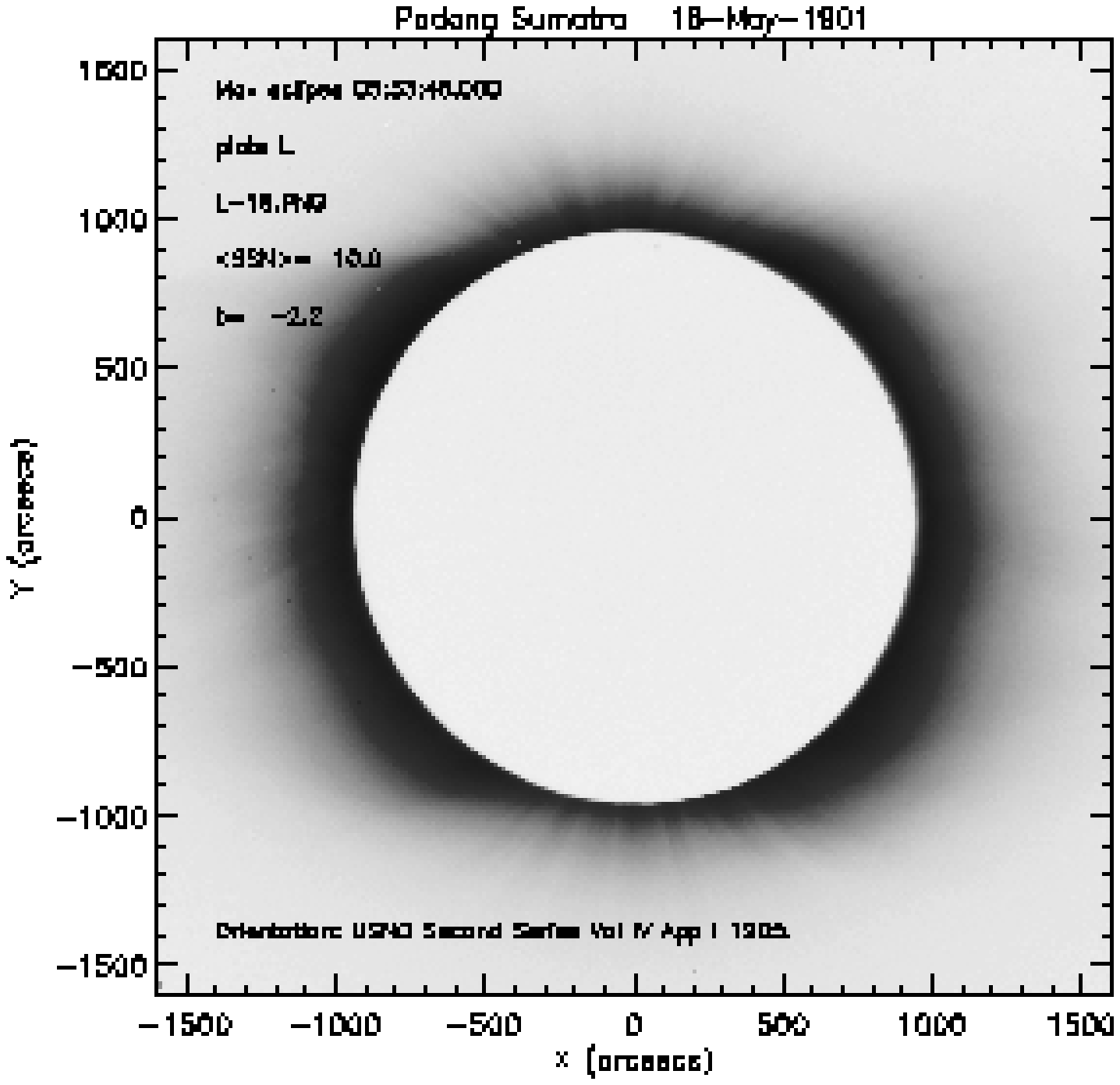}{2.6in}{0.}{50.}{50.}{-220}{-20}
\plotfiddle{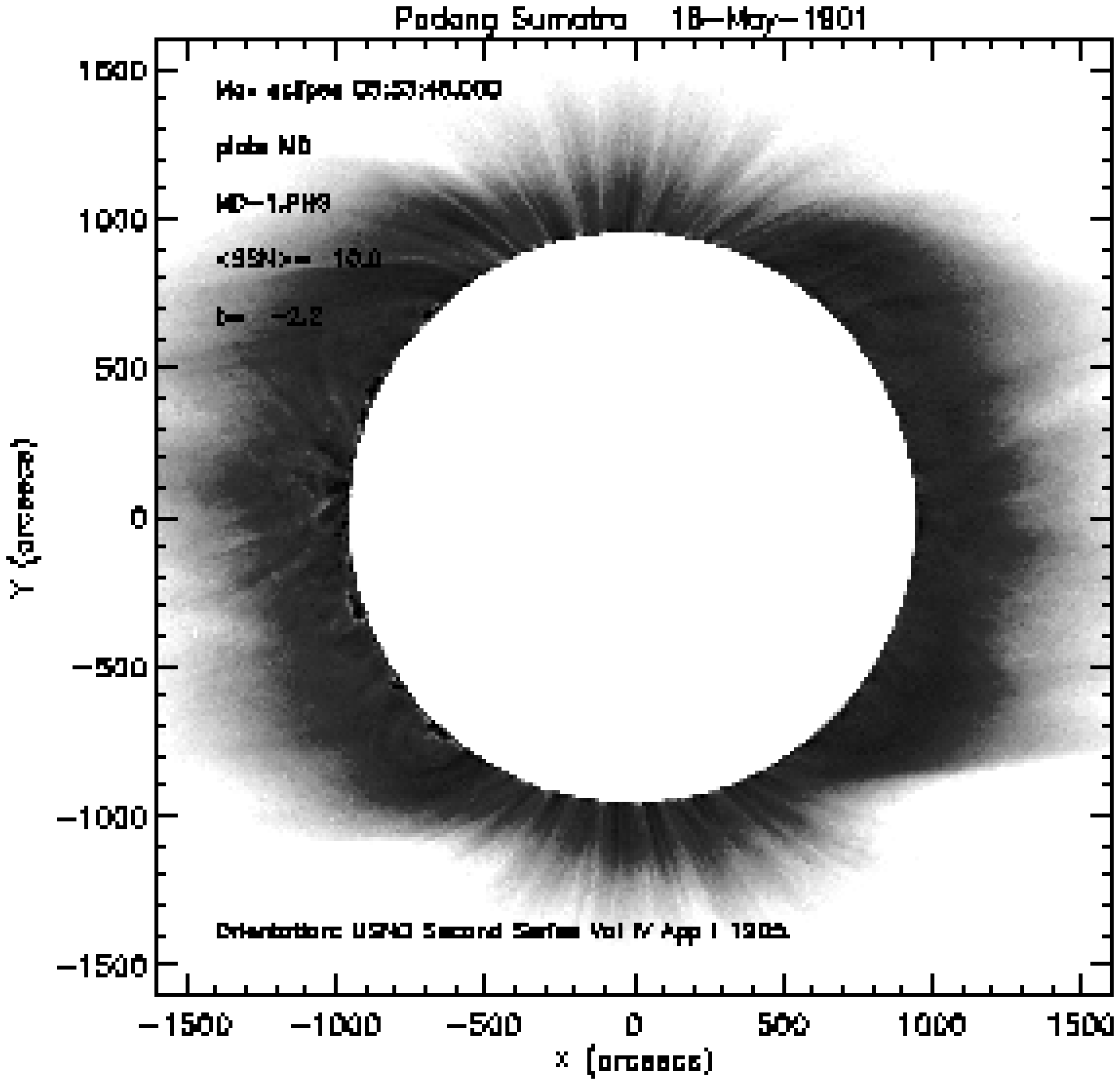}{2.6in}{0.}{50.}{50.}{-10}{180}
\vskip -2.5truein
 \caption{{\itshape Left:\/} One scanned negative plate for the
   eclipse of 1901.  
 {\itshape Right:\/} Aligned and filtered image which is a composite
 of several exposures.}
 \end{figure}

For comparison, we show in Figure 4 data obtained by one of us (MD)
using CCD imaging techniques, and processed in the same manner, for
the 2008 and 2009 eclipses.   Note that {\em only the morphology can really
be compared between the photographic and CCD data}, owing to
non-linearities inherent in photography, the filtering techniques, and
the color tables chosen simply to highlight the coronal structures.
We see that
the morphology of the 2008/2009 corona is qualitatively similar to 
the corona in 1901.  It is certainly not a ``canonical solar minimum''
corona consisting of a simple dipole with an equatorial current
sheet- such a configuration is represented by the image shown in
Figure~5. 
 Instead we see that there are prominence cavities around the
polar crown (see articles by Burkepile and Altrock in this volume).

 \begin{figure}[!h]
\plotfiddle{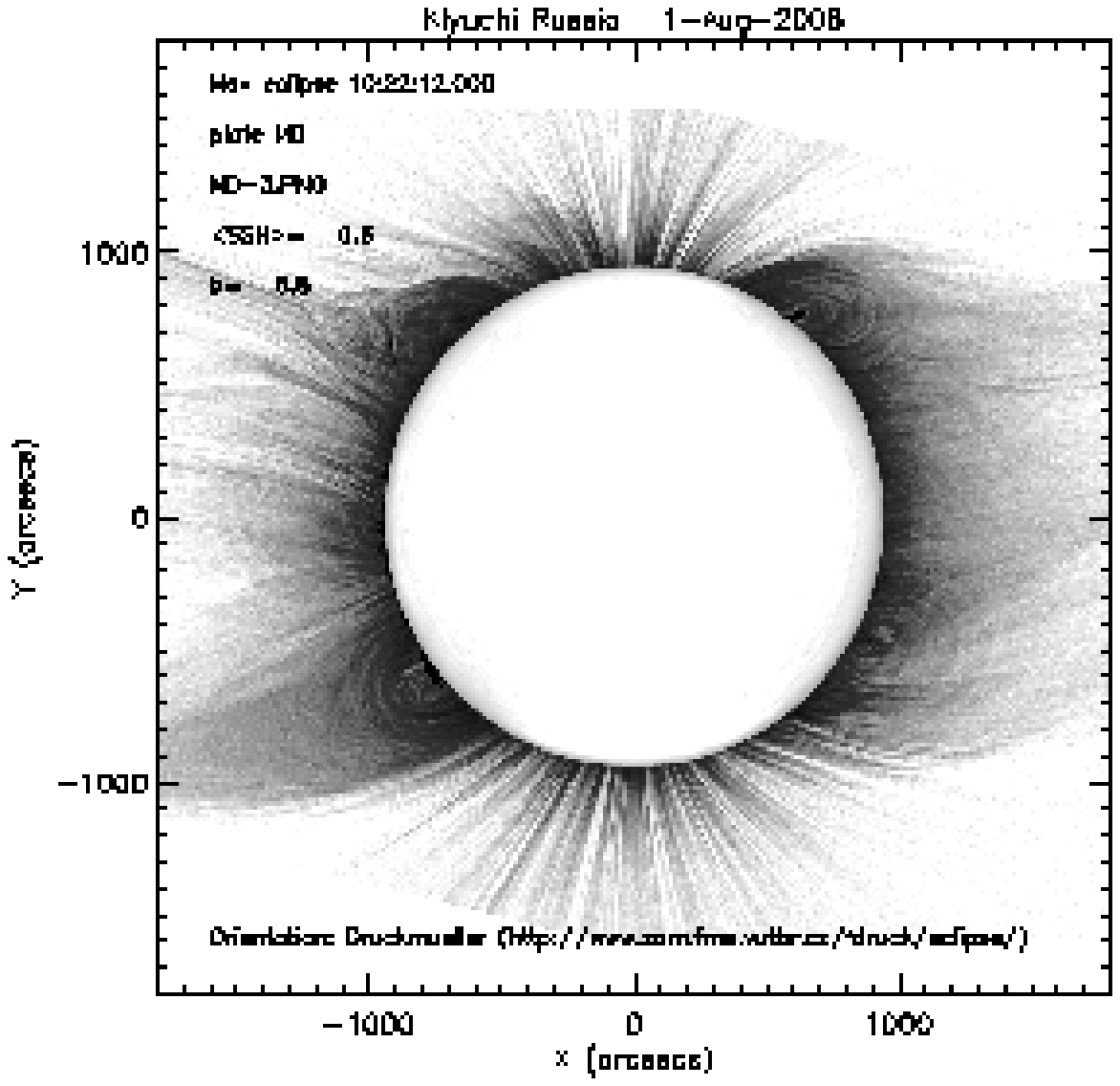}{2.6in}{0.}{50.}{50.}{-220}{-20}
\plotfiddle{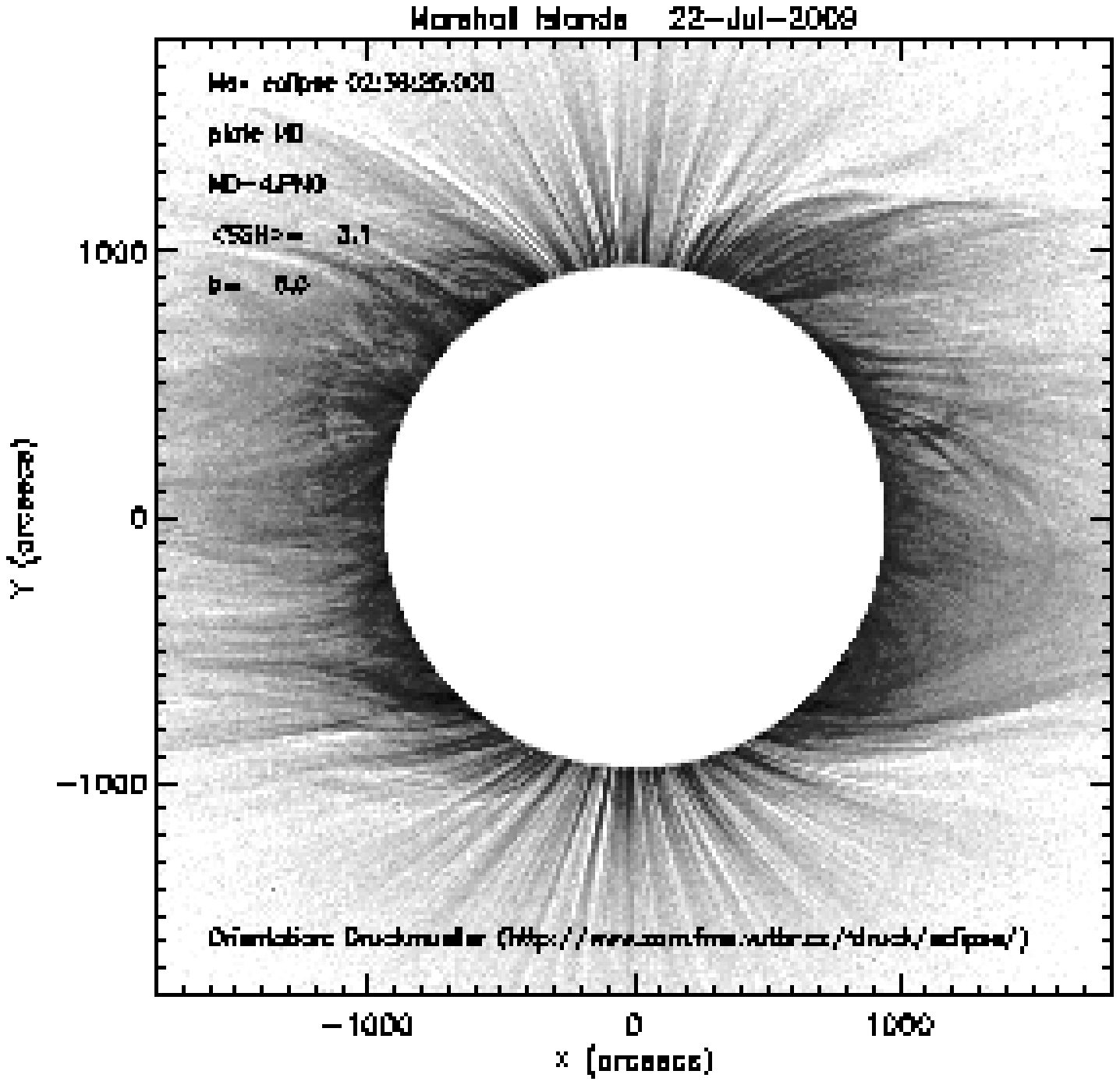}{2.6in}{0.}{50.}{50.}{-10}{180}
\vskip -2.5truein
 \caption{{\itshape Left:\/} Processed CCD image for the 2008
   eclipse, obtained and processed by Druckm\"uller.
{\itshape Right:\/} Similar data for the 2009 eclipse.}
 \end{figure}

\begin{figure}[!h]
 \plotone{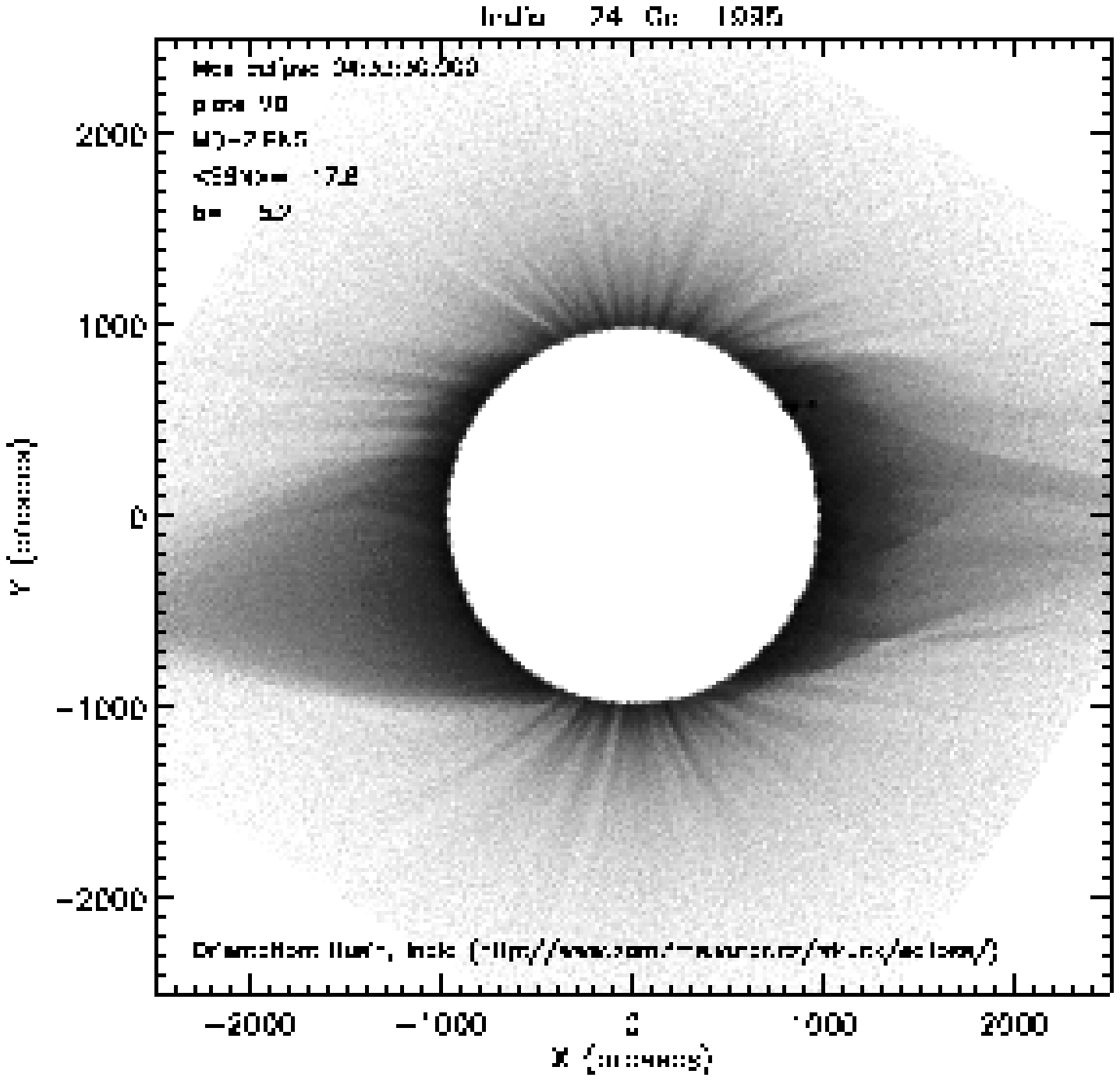}
 \caption{A 1995 eclipse image obtained by Rusin and processed by Druckm\"uller.} 
 \end{figure}

\vskip 18pt
\centerline{{\em Conclusions and speculations}}
\vskip 12pt

It is sometimes believed that more information on the coronal
morphology is contained in the 19th and early 20th century drawings of
eclipses than in the photographic records. But we have shown that the
historical eclipse plates, when digitized and carefully processed, can
approach a quality which enables meaningful comparison with modern CCD
images.  We have shown that the recent solar minimum corona has a
morphology very similar to eclipses from 1900 and 1901, and perhaps
1878, which is to be re-processed.  

Qualitative differences seem to exist between a ``classic''
dipolar/current sheet minimum configuration (Figure 5) and the
extended minima apparent in other figures.  Eclipse data for extended
minima show higher order multipolar structure, associated with high
latitude prominence cavities. In coronagraph data much of the
low-lying structure lies beneath the occulter.  In the recent minimum,
coronagraph data are dominated by two streamers either side of the
equator (Burkepile, this volume), structure also manifested as a ``warped'' heliospheric
current sheet, consistent with the prominence cavity structure
evident in Figure 4.

We speculate on two different causes of these different properties.  It is
common to invoke the strength and area of the unipolar polar fields as
a dominant influence on the coronal morphology.  But this can occur
mostly at heights $\gta 0.5R_\odot$, say, when the excess magnetic pressure
from strong polar fields 
pushes magnetic structures equator-ward, producing a classic
dipole/current sheet corona (Figure~5).  However, below $\sim 0.5R_\odot$,
local magnetic sources tend to dominate the magnetic force balance.
This is obviously more evident in eclipse data than in coronagraph
data, simply because of the angular size of the occulter.  We can
apply arguments concerning the distribution of magnetic polarity on
the Sun's surface put forward by \citet{Callebaut+Makarov+Tlatov2007}.
While these authors discuss non-simultaneous polar field reversals, we
draw attention to lower latitude regions.  During extended minima,
little flux emergence occurs at low latitudes, but emergence from the
new cycle begins.  Trans-equatorial coronal structures may then be
relatively dim.  At the same time meridional circulation continues to
transport the footpoints of the previously emerged magnetic flux
pole-ward.  In time, less magnetic flux crosses the equator, as coronal
magnetic connections are re-made at higher latitudes. The equatorial
current sheet weakens, and the coronal images appear more multipolar,
as seen in eclipse data for 2008, 2009 and 1901.

Perhaps the primary differences relate ultimately to the total
magnetic field generated by the dynamo.  If the total emerging
magnetic flux in cycle 23 were less than that in cycle 22, it would
take longer to reverse the polar fields, there would be more mixed
polarity at the poles, and weaker fields near $1R_\odot$.

The HAO eclipse archive will be made publically available 
soon\footnote{http://mlso.hao.ucar.edu/mlso\_eclipses.html}.
Readers are encouraged to contact HAO staff if they have 
data that might be added to this archive.

\end{document}